\documentclass[aps,prb,superscriptaddress,twocolumn,floatfix]{revtex4}
\usepackage{psfig}

\begin{document}

\title{
\parbox{30mm}{\fbox{\rule[1mm]{2mm}{-2mm}\Large\bf\sf PREPRINT}}
\hspace*{4mm}
\parbox{100mm}{\footnotesize\sf
Submitted for publication in {\it Optics Letters} (2005) }\hfill
\\[5mm]
Transmission of Slow Light through Photonic Crystal Waveguide Bends}
\date{\today}

\author{Solomon Assefa, Sharee J. McNab, and Yurii A. Vlasov}
\address {IBM T.~J.\ Watson Research Center, Yorktown Heights, NY 10598, USA}

\begin{abstract}
The spectral dependence of a bending loss of cascaded 60-degree bends in photonic crystal
(PhC) waveguides is explored in a slab-type silicon-on-insulator system. Ultra-low
bending loss of (0.05$\pm$0.03)dB/bend is measured at wavelengths corresponding to the
nearly dispersionless transmission regime. In contrast, the PhC bend is found to become
completely opaque for wavelengths range corresponding to the slow light regime. A general
strategy is presented and experimentally verified to optimize the bend design for
improved slow light transmission.
\end{abstract}

\maketitle

Planar two-dimensional photonic crystal (PhC) waveguides have a
strong potential to realize ultra-dense photonic
integrated-circuits. An important component of a compact photonic
circuit is a small-radius low-loss waveguide bend that efficiently
maneuvers light around sharp corners \cite{ref1}. Many papers have
investigated PhC bends focusing mainly on the optimization of the
amplitude transmission \cite{ref2-ref9}. Largely ignored, however,
is the spectral dependence of the bend loss. It is known that PhC
waveguides are characterized by very large group velocity dispersion
especially at wavelengths close to the onset of the waveguiding mode
\cite{ref10,ref11}. The strong distributed feedback at these
wavelengths results in significantly reduced group velocity. The
possibility to slow down the propagation of light is envisioned to
have a broad range of applications from all-optical data storage to
quantum computing \cite{ref12}. Therefore transmission in this "slow
light" regime needs to be carefully studied for various optical
components, if the "slow light"-based circuits are to be built. This
paper focuses on the spectral dependence of losses of PhC bends with
a special emphasis on the "slow light" regime.

The PhC waveguides were fabricated on 200mm silicon-on-insulator (SOI) wafers with 2$\mu$
m thick buried oxide (BOX) and 220nm thick silicon layer. The periodic array of holes
with a triangular lattice (lattice constant $a$=437nm) was first written in the resist
layer with electron beam lithography. The PhC waveguides were formed by removing a row of
holes in the $\Gamma$-K direction of the lattice. The pattern was then transferred to an
oxide hard mask using $CF_{4}$/$CHF_{3}$/Ar chemistry and the Si layer was etched next
with an HBr-based chemistry. The underlying BOX was wet etched in the PhC section to
create a suspended Si membrane. An additional lithography step defined the polymer
cladding of the inverted taper spot-size converter for efficient fiber coupling
\cite{ref13,ref14}.

\begin{figure}[tb]
\begin{center}
\leavevmode \psfig{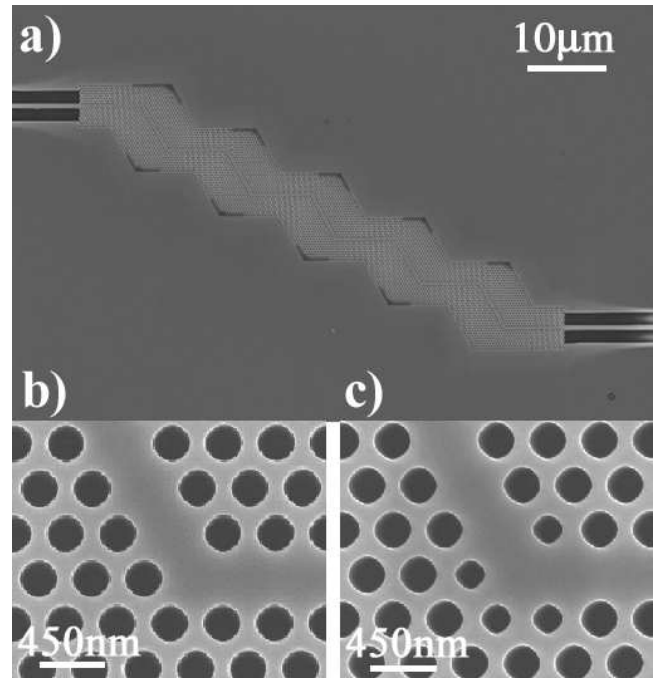}
\end{center}
\caption{Scanning electron micrographs of a PhC waveguide structures. A) 10 cascaded
60-degree bends, B) Magnified view of a simple 60-degree bend, C) Magnified view of a
modified 60-degree bend with four holes adjacent to the bend corner having smaller
radii.}
\label{fig1}
\end{figure}

In order to improve the experimental accuracy of loss measurements, devices with a number
of bends cascaded in series were fabricated as shown in Fig.1a. The bends were separated
from each other by straight waveguide sections 20 lattice constants long. To optically
characterize the devices, light from a broadband LED source (1200 to 1700nm) is coupled
to the PM fiber, transferred through the polarization controller and finally launched
into the device with a microlensed PM fiber tip. Transmission spectra were measured for
the TE polarization (E-field parallel to the slab plane) with a resolution of 5nm. The
measured transmission spectra through straight and bent PhC waveguides were normalized on
transmission through a reference straight strip waveguide following the method of Ref.
\cite{ref13}.

\begin{figure}[tb]
\begin{center}
\leavevmode \psfig{figure=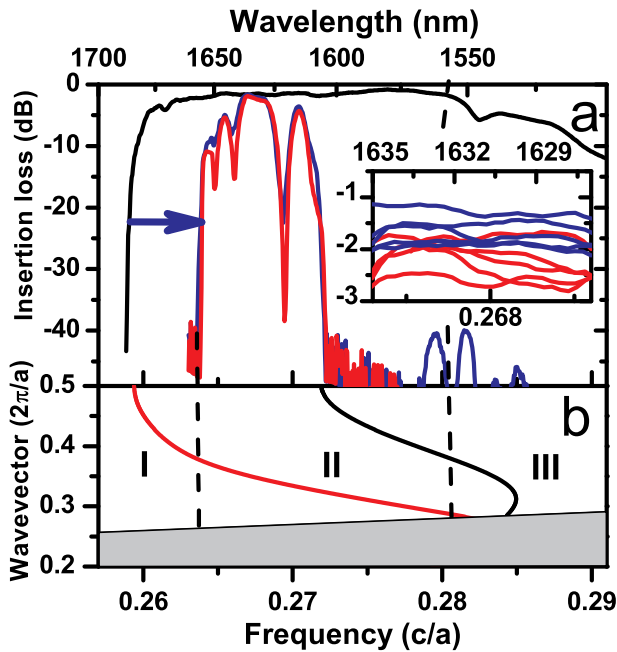,width=85mm}
\end{center}
\caption{Transmission spectra for PhC structures with hole radius of 100nm (0.23a).  A)
Spectrum (black curve) of a straight PhC waveguide with a length of 181$\mu$m and spectra
of structures with 10 (blue curve) and 20 (red curve) cascaded simple bends. Inset: Set
of spectra measured on ten different devices with 10 (red curves) and 20 (blue curves)
cascaded simple bends. B) Photonic band diagram for a PhC waveguide with a radius of
holes 0.23a and a slab thickness of 0.52a.}
\label{fig1}
\end{figure}

The design of the PhC bend similar to the one shown in Fig.1b was characterized first.
This simple bend is formed by connecting two W1 waveguides rotated by 60 degrees with
respect to each other. Figure 2 represents experimental transmission spectra of 10 and 20
of such cascaded simple bends formed in a PhC lattice with a hole radius of 100nm
(0.23a). The spectra show a transmission band with ~50nm bandwidth defined by two sharp
cut-offs around 1605 and 1655nm. Spectra measured for 10 and 20 cascaded bends are almost
identical, especially for 1627-1635nm spectral region shown in the inset in magnified
view, indicating low bending loss. Statistical analysis of measurements performed on a
number of identical devices gives a cumulative loss of 0.55dB per 10 bends that includes
a small loss component of 0.07dB which is the expected loss of a straight PhC waveguide
sections of equivalent length (90$\mu$m) \cite{ref14}. The standard deviation is
estimated for these measurements as 0.33dB, owing mainly to the small deviation in the
coupling efficiency from one device to another. This gives the bending loss per single
bend of (0.05$\pm$0.03)dB/bend. To our knowledge this is the lowest bending loss that has
yet been reported for PhC bends.

If bends are to be used to construct complex PhC-based circuits such as a Mach-Zehnder
interferometer employing straight PhC waveguides and multiple bends, transmission
characteristics of the bend need to be analyzed with respect to different transmission
regimes of a straight PhC waveguide. The black curve in Fig.2a represents the
transmission spectrum of a 181$\mu$m long straight PhC waveguide which corresponds to the
same length as the devices with 20 cascaded bends. Comparison of the experimental
spectrum with the photonic band structure calculated with the 3D plane-wave method
\cite{ref15} shown in Fig.2b allows identification of the sharp cut-off at 1687nm as the
onset of the waveguiding mode. The waveguide is characterized by very strong group
velocity dispersion with a group index becoming increasingly large in the region
1660nm-1687nm. This region can be defined as the "slow light" regime (region I in
Fig.2b). The dispersion is almost linear, corresponding to nearly constant group index,
in the wavelength region between 1660nm and 1560nm, which can be defined as the "linear"
regime (region II). At wavelengths above 1560nm (region III), where the mode is crossing
the light-line, the losses are increasing progressively due to out-of-plane leakage.

The transmission spectra of the bend exhibit drastically different
behavior for wavelengths corresponding to the three regimes. At
wavelengths corresponding to region III, where the mode is becoming
leaky for the straight PhC waveguide, the bend does not transmit any
light as shown in Fig.2a presumably due to significant increase in
out-of-plane losses at the bend. However, since the loss for a
straight PhC waveguide is already very large here, approaching
1500dB/cm \cite{ref13,ref14}, region III is not very useful for
integrated photonics.

In contrast, bending loss in the "linear" regime (region II) is
extremely small not exceeding 0.05dB/bend as demonstrated above.
This low bending loss is comparable to bending losses measured in
tightly curved bends in single-mode strip waveguides \cite{ref16}.
However, as opposed to strip waveguide bends operating over
broadband wavelengths, the PhC bend does not have a continuous
transmission band. In Fig.2a at wavelengths corresponding to region
II a series of sharp dips are seen with attenuation significantly
increasing as the number of cascaded bends increases, indicating
much higher bending loss of the order of 1dB/bend. These dips result
from small stopbands formed in the experimental optical circuit in
which straight and bend sections are repeated periodically.

Attaining low bending loss in the region I is crucial for
constructing optical circuits utilizing slow group velocity.
Surprisingly, comparison of transmission spectra (Fig.2a) of a
straight PhC waveguide and PhC bends reveals that bends are
completely opaque in this region. The long-wavelength cutoff of the
bend around 1655nm appears to be blue shifted by ~30nm (blue arrow)
with respect to the 1687nm cutoff characteristic of the straight PhC
waveguide. This blueshift is too large to be explained by small
variation of the lattice parameters between reference PhC straight
waveguide and the bend structure. Moreover, detailed inspection of a
number of SEM images as in Fig.1 reveals that the hole radii and the
lattice constant in both structures are identical within ±1nm.

\begin{figure}[tb]
\begin{center}
\leavevmode \psfig{figure=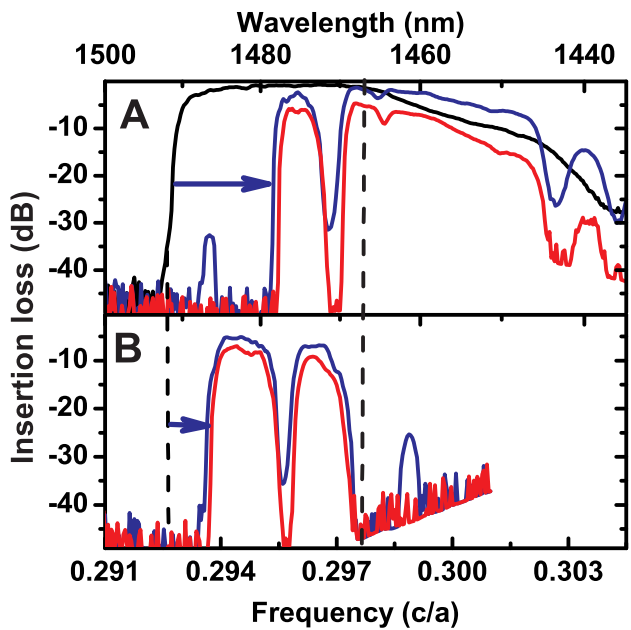,width=85mm}
\end{center}
\caption{Transmission spectra for PhC structures with a hole radius of 144nm (0.33a). A)
Spectrum (black curve) of a straight PhC waveguide with a length of 181um and spectra of
structures with 10 (blue curve) and 20 (red curve) cascaded simple bends. B) Spectra of
structures with 10 (blue curve) and 20 (red curve) cascaded modified bends. Arrows
indicate a blueshift of the cutoff.}
\label{fig1}
\end{figure}

This spectral blueshift can be explained by making analogy to
conventional strip and rib waveguide bends \cite{ref17}. It is known
that the mode in curved waveguides is pulled away from the center of
curvature toward the outer edge of the bend, thus seeing a lower
average refractive index with respect to the mode in the straight
waveguide \cite{ref17}. If this mechanism can be extended to
describe mode behavior in PhC waveguide bends, the PhC Bloch mode is
pulled toward its outer corner where it samples more holes. Hence,
the average refractive index that the mode encounter at the bend is
smaller, which explains the experimentally observed blueshift of the
cutoff. In integrated optics, the usual way to compensate the
distortion of the mode profile at the bend is to offset the bend
section with respect to the straight section or to increase the
refractive index at the bend by increasing its width \cite{ref17}.
Recent theoretical calculations have confirmed that a similar
hypothesis is applicable to PhC waveguide bends \cite{ref18}. A
modified design is proposed where the average refractive index at
the bend is effectively increased by decreasing the radius of holes
in the bend vicinity \cite{ref18}.

To experimentally verify this strategy, an additional set of devices was investigated,
fabricated on the same wafer as the previous devices. The hole radii of the PhC lattice
for this set was chosen to be 144nm (0.33a) in order to compare the experimental results
directly to the theoretical calculations of Ref.\cite{ref18}. This device set consisted
of cascaded simple bends analogous to that studied earlier (see Fig.1b), and the cascaded
modified bends (Fig.1c) designed for better transmission in the slow light regime
\cite{ref18}.

Figure 3a shows transmission spectra of a straight PhC waveguide and
a simple cascaded PhC bends. The low-loss transmission window for a
straight PhC waveguide is defined by a long-wavelength cutoff at
1493nm and a light-line at 1466nm. Band structure calculations
similar to Fig.2b showed that the "slow light" regime I in this case
corresponds to wavelengths range of 1478nm-1493nm. Similar to the
results obtained in Fig.2a, the spectrum of the simple bend is
characterized by a relatively low-loss transmission band extending
over the "linear" regime and interrupted by sharp dips. The
long-wavelength cutoff of a bend spectrum at 1478nm is blue shifted
by almost 15nm with respect to the cutoff position in a straight PhC
waveguide. Thus, the simple bend in this structure appears to be
completely opaque in the "slow light" regime as previously observed
in the results of Fig.2.

Figure 3b represents spectra of 10 and 20 cascaded bends, which are
modified by reducing the radius of four holes adjacent to the bend
section to 105nm (0.24a) as shown in Fig.1c. Transmission spectra of
10 and 20 cascaded modified bends have similar spectral features as
seen in the spectrum of a simple bend. The "linear" regime of a
modified bend is characterized by a relatively low loss ~0.1dB/bend
comparable to the results of Fig.2a.  As expected, the
long-wavelength cutoff of the modified bend occurs at wavelengths
longer than for a simple bend. Hence, by increasing the average
refractive index at the bend the initial 15nm blueshift of the
cutoff is significantly reduced to only 6nm rendering a large
portion of the "slow light" regime transparent.

The modified bends with 4 small holes clearly do not fully
compensate the initial blueshift. The experiments described above
can be viewed as a proof-of-principle demonstration of a general
strategy, rather than a final solution. Following the analogy with
the bends in conventional waveguides, the transmission in the "slow
light" regime might be improved not only by the increase of the
refractive index at the bend as shown here, but also by creating an
offset of the bend section with respect to the straight PhC
waveguide.

In conclusion, ultra-low bending loss of only (0.05$\pm$0.03)dB/turn is measured in a
simple 60-degree PhC bend. It is found that simple PhC bends are completely opaque for
the wavelength region corresponding to small group velocity regime of a straight PhC
waveguide. Increasing the average refractive index at the bend by decreasing the radius
of 4 holes adjacent to the bend section allows to almost compensate the blueshift of the
transmission cutoff and to restore transparency for slow light.

Discussions with Drs. N. Moll and G.-L. Bona are gratefully acknowledged. This work was
partially supported by DARPA/ONR (DSO, J. Lowell) through grant No.00014-04-C-0455.


\begin{thebibliography}{99}




\bibitem{ref1} A. Mekis, J.C. Chen, I. Kurkland, S. Fan, P. R. Villeneuve, and
J. D. Joannopoulos, "High Transmission through Sharp Bends in
Photonic Crystal Waveguides", Phys. Rev. Lett. 77, 3787 (1996).

\bibitem{ref2} T. Baba, N. Fukaya and J. Yonekura "Observation of light
propagation in photonic crystal optical waveguides with bends",
Electron. Lett. 35, 654 (1999)

\bibitem{ref3} M. Loncar, D. Nedeljkovic, T. Doll, J. Vuckovic, A. Scherer,
T.P. Pearsall, "Waveguiding in planar photonic crystal" , Appl.
Phys. Lett. 77 1937 (2000)

\bibitem{ref4} E. Chow, S. Y. Lin, J. R. Wendt, S. G. Johnson, J. D.
Joannopoulos, "Quantitative analysis of bending efficiency in photonic-crystal waveguide
bends at $\lambda$ = 1.55$\mu$m wavelengths," Opt. Lett. 26, 286-288 (2001).

\bibitem{ref5} Chutinan, M. Okano, S. Noda, "Wider bandwidth with high
transmission through waveguide bends in two-dimensional photonic
crystal slabs," Appl. Phys.Lett. 80, 1698 (2002)

\bibitem{ref6} L.H. Frandsen, A. Harpøth, P.I. Borel, M. Kristensen, J.S.
Jensen and O. Sigmund, "Broadband photonic crystal waveguide 60°
bend obtained utilizing topology optimization", Opt. Express 12,
5916 (2004).

\bibitem{ref7} I. Ntakis,P. Pottier, and R. M. De La Rue, "Optimization of
transmission properties of two-dimensional photonic crystal channel
waveguide bends through local lattice deformation", J. Appl. Phys.
96, 12 (2004)

\bibitem{ref8} B. Miao, C. Chen, S. Shi, J. Murakowski, and D. W. Prather,
"High-Efficiency Broad-Band Transmission Through a Double-60 Bend in
a Planar Photonic Crystal Single-Line Defect Waveguide", IEEE
Photon. Techn. Lett. 16 2469 (2004)

\bibitem{ref9} N. Ikeda, Y. Sugimoto, Y. Tanaka, K. Inoue, and K. Asakawa, "Low
Propagation Losses in Single-Line-Defect Photonic Crystal Waveguides
on GaAs Membrane" IEEE J. Sel. Areas Comm.  23, 1315 (2005).

\bibitem{ref10} M. Notomi, K. Yamada, A. Shinya, J. Takahashi, C. Takahashi, and
I. Yokohama "Extremely Large Group-Velocity Dispersion of
Line-Defect Waveguides in Photonic Crystal Slabs" Phys. Rev. Lett.
87, 253902 (2001)

\bibitem{ref11} Yu. A. Vlasov, M. O'Boyle, H. F. Hamann, S. J. McNab, "Taming
slowlight in photonic crystal waveguides" accepted for publication
(2005)

\bibitem{ref12} M.Yanik, W.Suh, Z.Wang, and S.Fan, "Stopping Light in a
Waveguide with an All-Optical Analog of Electromagnetically Induced
Transparency", Phys. Rev. Lett. 93, 233903 (2004)

\bibitem{ref13} S. J. McNab, N. Moll and Yu. A. Vlasov, "Ultra-low loss photonic
integrated circuit with membrane-type photonic crystal waveguides",
Opt. Express, 11, 2927 (2003)

\bibitem{ref14} E.Dulkeith, S.J. McNab and Yu.A. Vlasov "Mapping the optical
properties of slab-type two-dimensional photonic crystal
waveguides", Phys. Rev. B 72, 15 Sep (2005).

\bibitem{ref15} S. G. Johnson and J. D. Joannopoulos, "Block-iterative
frequency-domain methods for Maxwell's equations in a planewave
basis," Opt. Express 8, 173 (2001).

\bibitem{ref16} Y. A. Vlasov and S. J. McNab, "Losses in single-mode
silicon-on-insulator strip waveguides and bends," Opt. Express 12,
1622 (2004).

\bibitem{ref17} M. K. Smit, E. C. M. Pennings, H. Blok "A Normalized Approach to
the Design of Low-Loss Optical Waveguide Bends", Journ. Lightwave
Techn. 11, 1737 (1993)

\bibitem{ref18} N. Moll, G.-L. Bona, "Bend design for the low-group-velocity
mode in photonic crystal-slab waveguides", Appl. Phys. Lett. 85,
4322 (2004)

\end{thebibliography}
\end{document}